\documentclass[fleqn,a4paper,12pt]{article}
\usepackage{amssymb}
\usepackage{makeidx}
\usepackage{amsmath}
\usepackage{graphicx}
\usepackage{lscape}
\usepackage{a4}
\usepackage{epsfig}
\begin{document}

\thispagestyle{empty}
\newcommand{\al}{\alpha}
\newcommand{\bt}{\beta}
\newcommand{\s}{\sigma}
\newcommand{\lbd}{\lambda}
\newcommand{\vp}{\varphi}
\newcommand{\va}{\varepsilon}
\newcommand{\gm}{\gamma}
\newcommand{\G}{\Gamma}
\newcommand{\p}{\partial}
\newcommand{\om}{\omega}
\newcommand{\be} {\begin{equation}}
\newcommand{\lo}{\left(}
\newcommand{\ro} {\right)}
\newcommand{\ee} {\end{equation}}
\newcommand{\ba} {\begin{array}}
\newcommand{\ea} {\end{array}}
\newcommand{\ds}{\displaystyle}
\begin{center}
{\large\bf Group classification of systems of non-linear
reaction-diffusion equations with general diffusion matrix. I.
Generalized Ginzburg-Landau equations}
\end{center}
\vspace{2mm}
\begin{center}
  A. G. Nikitin\\

Institute of Mathematics of Nat.Acad. Sci of ukraine, 4
Tereshchenkivska str. 01601 Kyiv, Ukraine
\end{center}

 \vspace{2mm}

{\abstract {Group classification of the generalized complex
Ginzburg-Landau equations is presented. An approach to group
classification of systems of reaction-diffusion equations with
general diffusion matrix is formulated.}}

\section{ Introduction}

Group classification of differential equations is one of the
central problems of group analysis. It specifies non-equivalent
classes of equations and open the way to applications of symmetry
tools such as constructing and group generation of exact
solutions, separation of variables, etc. One of the goals of group
classification is a priori description of mathematical models with
a desired symmetry (e.q., relativistic invariance).

The first (and very impressive) achievements in group
classification belong to S. Lie who who had classified second
order ordinary differential equations and specified all cases when
such equations can be integrated in quadratures \cite{lie1}. Lie
had presented also a group classification of an entire class of
partial differential equations, namely, linear equations
with two independent variables. In particular, it was Lie who for
the first time describes group properties of the linear heat
equation \cite{lie2}.

The next step in classification of heat equations was made by
Dorodnitsyn \cite{OV} who had classified nonlinear diffusion equations
\be u_t -u_{xx}=f(u)\label{d}\ee where $f$ is a function of
$u=u(t,x)$ and  subscripts denote derivations w.r.t. the
corresponding variables \footnote{In paper \cite{OV} a more general equations with nonlinear diffusion were
classified which include (\ref{d}) as a particular case}
. This result was extended by Fushchych ,
Serov~ \cite{FU} and Clarkson and Mansfield~ \cite{CL} to the case
of  non-classical (conditional) symmetries.

The results of group classification of equations (\ref{d}) play an
important role in constructing of their exact solutions and
qualitative analysis of the nonlinear heat equation, refer, e.g.
to \cite{sam}.

In the present paper we perform the group
classification of systems of the nonlinear reaction-diffusion
equations \be \label{1.1}\ba{l}
\displaystyle u_t-\Delta(a u- v)=f^1(u,v),\\
\displaystyle v_ t-\Delta( u+a v)=f^2(u,v) \ea \ee
where $u$ and $v$ are function of $t, x_1, x_2,  \ldots , x_m$,
$a$ is a real constant
and $\Delta$ is the Laplace operator in $R^m$. We shall write
(\ref{1.1}) also in the matrix form:
     \be \label{1.2} U_t- A\Delta U=f(U) \ee where $A$ is a matrix whose elements are
$A^{11}=A^{22}=a, \ A^{12}=-A^{21}=b$, $U=\left(\ba{l}u\\v\ea\ro$ and $f=\left(\ba{l} f^1 \\
f^2\ea \right)$.

Mathematical models based on equations (\ref{1.1}) are widely used
in mathematical physics, biology, chemistry, etc. Here
we present only two significant examples.

\begin{itemize}
\item The nonlinear Schr{\"{o}}dinger (NS) equation  in
$m-$dimensional space:
\begin{equation}
\left( i\partial_ t+\Delta\right) \psi =F(\psi
,\psi ^{*}) \label{d6}
\end{equation}
is a particular case of (\ref{1.1}). If we denote $ \psi=u+iv$,
$F=f_1+if_2$ then (\ref{d6}) reduces to the form (\ref{1.2}) with
$A=\lo\ba{rr}0&-1\\1&0\ea\ro$.

Equations (\ref{d6}) with various nonlinearities $F$ are used in
nonlinear optics, non-linear quantum mechanics \cite{doebner},
they serve as one of basic models of inverse scattering problem
\cite{faddeev}. The most popular models are connected with the
following nonlinearities \cite{FU1}:
\[
F=F(\psi ^{*}\psi )\psi,\quad F=(\psi ^{*}\psi )^k\psi,\qquad
F=(\psi ^{*}\psi )^{\frac 2m}\psi,\quad F=\ln (\psi ^{*}\psi )\psi
\] One more interesting particular case of the NS equation corresponds to
\[\left( i{\partial_ t}+\Delta\right) \psi =(\psi-\psi^*)^2;\]
in this case (\ref{d6}) is a potential equation for the Boussinesq
equation for function $V={\p_ t}(\psi-\psi^*)$.

Group classification of the NS equation has been performed in paper \cite{pop}.

\item Generalized complex Ginzburg-Landau (CGL) equation
\be W_\tau-(1+i\beta)\Delta W=F(W,W^*)\label{la}\ee
also can be treated as a particular case of system (\ref{1.2}).
Indeed, representing $W$ and $F$ as $W=(u+iv), F=\beta(f^1+if^2)$
and changing independent variable $\tau \to t=\beta\tau$ we
transform (\ref{la}) to the form (\ref{1.2}) with
$A=\lo\ba{ll}\beta^{-1}&-1\\1&\beta^{-1}\ea\ro$. The standard CGL
equation corresponds to the case $F=W-(1+i\alpha)W|W|^2.$
\end{itemize}

Thus the symmetry analysis of equations (\ref{1.1}) has a large
application value and can be used, e.g., to construct exact
solutions for a very extended class of physical and biological
systems. The comprehensive group analysis of systems (\ref{1.1})
is also a nice "internal" problem of the Lie theory which admits
exact general solution for the case of {\it arbitrary} number of
independent variables $x_1, x_2, \ldots , x_m$.

We notice that group classification of equations (\ref{1.1}) by no
means is a standard problem of group analysis of partial differential equations
 which can be
solved with direct application of well-known algorithms. Because
of presence of two arbitrary elements, i.e., $f^1$ and $f^2$, this
classification needs a rather nontrivial generalization of the
approach \cite{OV} used for classification of equation (\ref{d}).

Equations (\ref{1.2}) with arbitrary invertible matrix $A$ were
classified in paper \cite{nikwil1}. To our great a pity, mainly due
to typographical errors (made during the editing procedure),
presentation of results in \cite{nikwil1} was not satisfactory
\footnote{The tables presenting the results of group classification
have been deformed and cut off. It is necessary to stress that it
was the authors fault, one of whom signed the paper proofs without
careful reading.}.

The present paper is the first from the series in which we present
the completed group classification of coupled reaction-diffusion
equations (\ref{1.2}) with {\it arbitrary} (i.e., invertible or singular) matrix $A$. Moreover,
we present a straightforward and easily verified procedure of
solution of the determining equations which guarantees the
completeness of the obtained results.

We also  indicate clearly the equivalence relations used in the
classification procedure, i.e., present explicitly the equivalence
groups for for all classified equations. In addition, we extend
the results obtained in \cite{nikwil1} to the important case of
{\it non-invertible} matrix $A$ and more general equations
including both the first and second order derivatives with respect to spatial variables.

Let us note that there are three ad hoc non-equivalent classes of
equations (\ref{1.2}) corresponding to the following forms of
matrices $A$

\be\label{2.1} \ba{l} I. \quad A=\left(\ba{cc} a & -1 \\ 1 & a \ea
\right);\quad II.\quad  A=\left(\ba{cc} 1 & 0 \\ 0 & a \ea
\right); \quad   \ III. \quad A=\left(\ba{cc} a & 0 \\ 1 & a \ea
\right)\ea \ee where $a$ is an arbitrary parameter. Moreover, any
$ 2\times 2$ matrix $A$ can be reduced to one of the forms
(\ref{2.1}) using linear transformations of dependent variables
and scaling independent variables in (\ref{1.2}).

The NS and CGL equations correspond to matrices $A$ of form $I$.
The general equations (\ref{1.2}) with such matrices (i.e.,
generalized CGL equations) are the main subject of group
classification carried out in the present paper while the cases
$II$ and $III$ will be considered in the following publications.
Nevertheless till an appropriate
moment we will consider equations with all types of matrices $A$
enumerated in (\ref{2.1}).

\section{Determining equations and equivalence transformations}

In the first stage we restrict ourselves to group classification
of equations (\ref{1.2}) with invertible matrix $A$. Moreover, till an appropriate moment
we consider
equations (\ref{1.2}) with arbitrary number $n$ of dependent variables.

  Using the
standard Lie algorithm \cite{olver} (or its specific version
proposed in \cite{nikwil1}) one can find determining equations for
the functions $\eta ,\xi^{a}$ and $\pi^{a}$ which specify
generator $X$ of the symmetry group admitted by equation
(\ref{1.2}):
\begin{equation}
X=\eta {\frac{\partial }{\partial t}}+\xi ^{\nu}{\frac{\partial
}{\partial x_{\nu}}}-\pi^{b}{\frac{\partial }{\partial
u_{b}}}\equiv \eta\p_t+\xi^\nu\p_{x_\nu}-\pi^b\p_{u_b} \label{3.105}
\end{equation}
where a summation from $1$ to $m$ and from $1$ to $2$ is assumed
over repeated indices $\nu$ and $b$ respectively, and a temporary
notation $u=u_1, v=u_2$ is used. In a more general case of $n$ dependent variables
$U=column(u_1,u_2,\cdots,u_n)$ the repeated indices $b$ run over the values $1,2,\cdots, n$.

We shall not reproduce the deduction of the determining equations
here (refer to \cite{nikwil1}) but present them directly.

Dependence of $\eta, \xi^\nu$ and $\pi^b$ on $U$ is defined by the following relations:
\begin{equation}
 \eta_{ u_{a}}=0,\ \xi ^{\nu}_{u_{b}}=0,\ \ \pi ^{a}_{u_{c}u_{b}}=0. \label{3.4a}
\end{equation}
So from (\ref{3.4a}) $\eta $ and $\xi ^{\nu}$ are functions of $t$ and $x_{\mu}$
and, $\pi ^{\nu}$ is linear in $u_{a}$. Thus:
\begin{equation}
\pi ^{a}=N^{ab}u_{b}+B ^{a}  \label{3.5a}
\end{equation}
where $N^{ab}, B^a$ are functions of $t $ and $x_{\nu}$ only.
The remaining equations are \cite{nikwil1}:
\begin{equation}
2A\xi _{x_\mu}^{\nu}=-\delta ^{\mu\nu}(\eta_t A+[A,N ]),\qquad {\eta}
_{x_\nu t}=0, \label{4.2a}
\end{equation}
\begin{equation}\ba{l}
{\xi}^\nu_t-2AN_{x_\nu}-A\Delta\xi^{\nu}=0, \\

\eta_tf^k+N^{kb}f^{b}+(N^{kb}_t-\Delta A^{ks}N^{sb})u_b
+B^k_t-\Delta A^{kc}B^c
=(B ^{a}+N^{ab}u_{b})f^{k}_{u_a}.\ea
\label{3.8a}
\end{equation}
Here $N$ and $A$ are matrices whose elements are $N^{ab}$ and $A^{ab}$, $\delta^{ab}$ is the
Kronecker symbol.

In accordance with (\ref{3.5a})--(\ref{3.8a})
the general form of the related generator
(\ref{3.105}) is \cite{nikwil1}: \be \label{2.4} \ba{l}
X=\lbd K+\sigma_\mu G_\mu+\om_\mu \hat G_\mu+\mu
D-(C^{ab}u_b+B^a){\p_{ u_a}}\\
+\Psi^{\mu \nu} x_\mu \p_{x_\nu}+\nu \p_t+\rho_\mu\p_{x_\mu} \ea
\ee where the Greek letters denote arbitrary constants, $B^a$ are
functions of $t,x$, and $C^{ab}$ are functions of $t$ satisfying
\be \label{2.5} C^{ab}A^{bk}-A^{ab}C^{bk}=0 \ee and \be
\label{2.6} \ba{l} K=2t(t\p_t+x_\mu
\p_{x_\mu})-\frac{x^2}{2}(A^{-1})^{ab}
u_b\p_{u_a}-tmu_a\p_{u_a}
,\\
G_\mu=t\p_{x_\mu}+\frac{1}{2}x_\mu(A^{-1})^{ab}u_b\p_{u_a},\\
\hat G_\mu=e^{\gm t}\left(\p_{x_\mu}+\frac{1}{2}\gm
x_\mu(A^{-1})^{ab}
u_b {\p}_{u_a}\ro,\\
D=t\p_t+\frac12 x_\mu \p_{x_\mu}. \ea \ee
    Here $A^{ab}$ and $(A^{-1})^{ab}$ are elements of matrix $A$ and
    matrix inverse to $A$ respectively.

In accordance with (\ref{3.8a}) equation (\ref{1.2}) admits symmetry operator (\ref{2.4})
 iff the following classifying equations for $f^1$ and
$f^2$ are satisfied:
    \be \label{2.7} \ba{l} (\lbd t(m+4)
+\mu)f^a+\left(\frac{\lbd}{2} x^2+\sigma_\mu x_\mu+\gm e^{\gm t}
\om_\mu x_\mu\right) (A^{-1})^{ab}f^b\\
+C^{ab}f^b+C^{ab}_t u_b +B_t^a-\Delta A^{ab}B^b\\
=(B^s+C^{sb} u_b+\lbd mtu_s +\left(\frac{\lbd}2 x^2+\sigma_\mu x_\mu+\gm e^{\gm t}\om_\mu
x_\mu\right) (A^{-1})^{sk}u_k) f^a_{u_s}. \ea \ee
Thus the group classification of equations (\ref{1.2}) with a
non-singular matrix $A$ reduces to solving equation (\ref{2.7})
where $\lbd, \mu, \sigma_\mu, \om_\mu, \gm $ are arbitrary
parameters, $B^a$ and $C^{ab}$ are functions of $(t,x)$ and $t$
respectively. Moreover, matrix $C$ with elements $C^{ab}$ should
commute with $A$.

We notice that relations (\ref{2.4})-(\ref{2.7}) are valid for
group classification of  systems (\ref{1.2}) of coupled
reaction-diffusion equations including {\it arbitrary} number $n$
of dependent variables $U=(u_1, u_2, \ldots u_n)$ provided the
related $n \times n$ matrix $A$ be invertible \cite{nikwil1}. In
this case indices $a, b, s, k$ in (\ref{2.4})-(\ref{2.7}) run over
the values $1,2 \ldots n$.

We will solve classifying equations (\ref{2.7}) up to equivalence
transformations $ U \to \tilde U=G(U,t,x)$, $t \to \tilde
t=T(U,t,x)$, $x \to \tilde x=X(U,t,x)$ and $ f \to \tilde
f=F(U,t,x,f)$ which keep the general form of equations (\ref{1.2})
but can change functions $f^1$ and $f^2$. The group of equivalence
transformations for equation (\ref{1.2}) can be found using the
classical Lie approach and treating $f^1$ and $f^2$ as additional
dependent variables. In addition to the obvious symmetry
transformations \be \label{2.2} t \to t'=t+a, \quad x_\mu \to
x'_\mu=R_{\mu \nu} x_\nu +b_\mu \ee where $a, b_\mu$ and $R_{\mu
\nu}$ are arbitrary parameters satisfying $R_{\mu \nu} R_{\mu
\lbd}=\delta_{\mu\lbd}$, this group includes the following
transformations \be \label{x.2} \ba{l}
u_a\to K^{ab}u_b+b_a, \quad f^a \to \lbd^2 K^{ab}f^b,\\
t \to \lbd^{-2} t, \quad x_a \to \lbd^{-1}x_a \ea \ee where
$K^{ab}$ are elements of an invertible constant matrix $K$
commuting with $A$, $\lbd \not=0$ and $b_a$ are arbitrary
constants.

For the case when $n=2$ and matrix $A$ belongs to type $I$ (\ref{2.1}) the
transformation matrix has the following form
    \be \label{x.5}
  K=\left(\ba{cc} K_{1} & - K_{2}\cr K_{2} & K_{1}\ea
\right), \quad K^2_1 + K^2_2  \not = 0. \ee

It is possible to show that there is no more extended equivalence
relations valid for arbitrary nonlinearities $f^1$ and $f^2$.
However, if functions $f^1, f^2$ are specified, the invariance
group can be more extended. In addition to transformations
(\ref{x.2}) it includes symmetry transformations generated by
infinitesimal operator (\ref{2.4}), and can include additional
equivalence transformations (AET). We will specify AET in the
following.

\section{Basic, main and extended symmetries}

Thus to describe Lie symmetries of equation (\ref{1.2}) (whose
generators have the general form (\ref{2.4})) it is necessary to
find all non-linearities $f^1$ and $f^2$ which satisfy the
corresponding classifying equations (\ref{2.7}). To solve these
rather complicated equations we use the main algebraic property of
the related symmetries, i.e., the fact that they should form a Lie
algebra. In other words, instead of going throw all non-equivalent
possibilities arising via separation of variables in the
classifying equations we first specify all non-equivalent
realizations of the invariance algebra for our equations. Then
using the one-to-one correspondence between these algebras and
classifying equations (\ref{2.7}) we easily solve the group
classification problems for equations (\ref{1.2}).

Equation (\ref{2.7}) does not include parameters $\Psi^{\mu \nu},
\nu$ and $\rho_\nu$ present in (\ref{2.4}) thus for any $f^1$ and
$f^2$ equation (\ref{1.2}) admits symmetries generated by the
following operators \be \label{4.1} P_0=\p_t, \quad
P_\lbd=\p_{x_\lbd}, \quad J_{\mu \nu}=x_\mu\p_{x_\nu}-x_\nu
\p_{x_\mu}. \ee

Infinitesimal operators (\ref{4.1}) generate the evident symmetry
transformations (\ref{2.2}) which form the kernel of invariance
groups of equation (\ref{1.2}). For some classes of nonlinearities
$f^1$ and $f^2$ the invariance algebra of equation (\ref{1.2}) is
more extended but includes (\ref{4.1}) as a subalgebra. We will
refer to (\ref{4.1}) as to {\it basic symmetries}.

Let us specify one more subclass of symmetries of equation
(\ref{1.2}) which we call {\it main symmetries}. The related
generator $\tilde X$ has the form (\ref{2.4}) with
$\Psi^{\mu\nu}=\nu=\rho_\nu=\sigma_\nu=\om_\nu=0$, i.e., \be
\label{4.2} \tilde X=\mu D+C^{ab}u_b \p_{
u_a}+B^a\p_{ u_a}. \ee

The classifying equation for symmetries
(\ref{4.2}) can be obtained from (\ref{2.7}) by setting
$\mu=\sigma^a=\om^a=0$. As a result we obtain \be \label{4.3} (\mu
\delta^{ab}+C^{ab})f^b+C^{ab}_t u_b+B^a_t-\Delta A^{ab}B^b=
(C^{nb}u_b+B^n)f^a_{u_n}. \ee

It is easily verified that operators (\ref{4.2}) and (\ref{4.1})
form a Lie algebra which is a subalgebra of symmetries for
equation (\ref{1.2}) (this algebra can be either finite of
infinite dimensional).

On the other hand, if equation (\ref{1.2}) admits a more general
symmetry (\ref{2.4}) with $\s_a\neq 0$ or (and) $\lambda\neq 0,\
\omega^\mu\neq 0$ then it has to admit symmetry (\ref{4.2}) also. To
prove this we calculate multiple commutators of (\ref{2.4}) with the
basic symmetries (\ref{4.1}) and use the fact that such commutators
have to belong to symmetries of equation (\ref{1.2}), i.e., generate
their own classifying equation (\ref{2.7}).

Let equation (\ref{1.2}) admits symmetry (\ref{2.4}) with
$\sigma_\mu\not=0, \Psi^{\mu\nu}=\rho_\mu=\nu=\lambda=\om^k=0$,
i.e., \be \label{4.4} X=\sigma_\nu G_\nu+\mu
D+(C^{ab}u_b+B^a)\p_{ u_b}. \ee

Commuting $Y$ with $P_\mu$ we obtain one more symmetry \be
\label{4.5} Y_\mu=-\frac{\sigma_\mu}{2}(A^{-1})^{ab} u_b
\p_{ u_a}+B^a_\mu \p_{ u_a}+\mu P_\mu. \ee

The last term belongs to the basic symmetry algebra (\ref{4.1})
and so can be omitted. The remaining terms are of the type
(\ref{4.2}).

Thus supposing the extended symmetry (\ref{4.4}) is admissible we
conclude that equation (\ref{1.2}) has to admit the main symmetry
also.

Commuting (\ref{4.5}) with $P_0$ and $P_\lbd$ we come to the
following symmetries: \be \label{4.6}
Y_{\mu\nu}=B^a_{\mu\nu}\p_{ u_a}, \ Y_{\mu t}=B^a_{\mu
t}\p_{ u_a}. \ee

Any symmetry (\ref{4.4})-(\ref{4.6}) generates this own system
(\ref{2.7}) of classifying equations. After strait forward but
rather cumbersome calculations we conclude that all these systems
are compatible provided the following condition is satisfied \be
\label{4.8} (A^{-1})^{ab}f^b=(A^{-1})^{nb}u_b f^a_{ u_n}. \ee

If (\ref{4.8}) is satisfied equation (\ref{1.2}) admits symmetry
(\ref{4.4}) with $\mu=C^{ab}=B^a=0$, i.e., Galilei generators
$G_\nu$ of (\ref{2.6}).

Analogously, supposing that equation (\ref{1.2}) admits
extended symmetry (\ref{2.4}) with $\lbd \not=0$
and $\om^a=0$ we conclude that it has to admit symmetry
(\ref{4.4}) with $\mu\not=0$ and $\sigma_\nu\not=0$ also. The related
functions $f^1$ and $f^2$ should satisfy relations (\ref{4.8}) and
(\ref{4.3}). Moreover, analyzing possible dependence of $C^{ab}$ and $B^a$
in the corresponding relations (\ref{2.7}) on $t$ we conclude that they
 should be ether scalars or linear in $t$, i.e.,
$ C^{ab}=\mu^{ab}t+\nu^{ab}$.
Moreover, up to equivalence transformations (\ref{x.2}) we can choose $ B^a=0$
and reduce the related equation system
(\ref{2.7}), (\ref{4.3}) to the following equations:
\be\label{!!}\ba{l}(m+4)f^a+\mu^{ab}f^b=(\mu^{kb}u_b+mu_k)\frac{\p f^a}{\p u_k},
\\\nu^{ab}f^b+\mu^{ab}u_b=\nu^{kb}u_b\frac{\p f^a}{\p u_k}\ea\ee
where constants $\nu^{ab}$ and $\mu^{ab}$ are non-trivial in the case
of the diagonal diffusion matrix only.

Finally for general symmetry (\ref{2.4}) it is not difficult to
show that the condition $\om^a\not=0$ leads to the following
equation for $f^a$ \be \label{4.10} (A^{-1})^{kb}(f^b+\gm
u^b)=(A^{-1})^{ab}u_b  f^k_{u_a}. \ee

We notice that relations (\ref{4.8}) and (\ref{4.10}) are
particular cases of (\ref{4.3}) for $ \mu=0 $,
$C^{ab}=(A^{-1})^{ab}$ and $ \mu=0 $, $C^{ab}=e^{\gm
t}(A^{-1})^{ab}$ respectively. Thus if relation (\ref{4.8}) is
valid then, in addition to $G_\mu$ (\ref{2.6}) equation
(\ref{1.2}) admits the symmetry \be \label{4.11}
X=(A^{-1})^{ab}u_b \p_{ u_a}. \ee

Alternatively, if (\ref{4.10}) is satisfied, equation (\ref{1.1})
admits symmetry $\hat G_\mu$ (2.6) and also the following one \be
\label{4.12} X=e^{\gm t}(A^{-1})^{ab}u_b \p_{ u_a}, \quad
\gm \not =0 .\ee

Thus it is reasonable first to classify equations (\ref{1.2})
which admit main symmetries (\ref{4.2}) and then specify all cases
when these symmetries can be extended.

The conditions when system (\ref{1.2}) admits extended symmetries
are given by relations (\ref{4.8})-(\ref{4.10}).

Results of Section 3 are valid for equations (\ref{1.2}) with
arbitrary invertible diffusion matrix. In the following we
restrict ourselves to the case when $n=2$ and matrix $A$ has the form $I$
given in (\ref{2.1}), i.e., to the case of generalized CLG
equations. In other words, we will classify the following systems
of coupled reaction-diffusion equations:
    \be
\label{LG}\ba{l}
\displaystyle u_t-\Delta(a u- v)=f^1(u,v),\\
\displaystyle v_ t-\Delta(a v +u)=f^2(u,v) \ea \ee which are
particular cases of general systems (\ref{1.2}) when matrix $A$ has the form $I$  (\ref{2.1}).

\section{ Algebras of main symmetries}

In accordance with the plane outlined above we start with
investigation of main symmetries (\ref{4.2}) admitted by equation
(\ref{LG}).

The first step of our analysis is to describe non-equivalent Lie
algebras $\cal A$ of operators (\ref{4.2}) which can be admitted
by this equation. We shall consider consequently one-, two-,...,
n-dimensional algebras $\cal A$.

For any type of matrix $A$ enumerated in (\ref{2.1}) we specify
all non-equivalent "tails" of operators (\ref{4.2}), i.e., the
terms
    \be \label{8.1}
N=C^{ab}u_b\p_{ u_a} + B^a \p_{ u_a}. \ee

These terms can either be a constituent part of a more general
symmetry (\ref{4.2}) or represent a particular case of (\ref{4.2})
corresponding to $\mu=0$. Thus the problem of classification of
algebras $\cal A$ includes a subproblem of classification of
algebras of operators (\ref{8.1}).

Let equation (\ref{1.2}) admits a one-dimensional invariance
algebra whose basis element have the form (\ref{8.1}), and does
not admit a more extended algebra of the main symmetries. Then
commutators of $N$ with the basic symmetries $P_0$ and $P_a$
should be equal to a linear combination of $N$ and operators
(\ref{4.1}). It is easily verified that there are three
possibilities: \be\ba{l} \label{8.4} 1.\ C^{ab}=\mu^{ab}, \quad
B^a=\mu^a,\\  2.\ C^{ab}=e^{\lbd t} \mu^{ab}, \quad B^a=e^{\lbd t}
\mu^a, \\ 3.\  C^{ab}=0, \quad B^a= e^{\lbd t+\om \cdot x} \mu^a
\ea\ee where $\mu^{ab}, \mu^a, \lbd $, and $\om$ are constants,
and matrix with elements $\mu^{ab}$ should commute with $A$. In
the case when $A$ is of the form (\ref{matrix})
constants $\mu^{ab}$ are restricted by the following relations:
$\mu^{11}=\mu^{22},\ \mu^{12}=-\mu^{21}$.

 To specify all non-equivalent operators (\ref{8.1}), (\ref{8.4})
 we use the isomorphism of (\ref{8.1}) with $3 \times 3$ matrices
 of the following form
    \be
\label{8.5} g=\left( \ba{ccc}
0   &  0       &  0\\
B^1 &  C^{11}  &  C^{12}\\
B^2 &  C^{21}  &  C^{12}  \ea \right) \sim \left( \ba{ccc}
0   &  0       &  0\\
\mu^1 &  \mu^{11}  &  \mu^{12}\\
\mu^2 &  -\mu^{12}  &  \mu^{11}  \ea \right). \ee

Equations (\ref{LG}) admit equivalence transformations (\ref{x.2})
which change the term $N$ (\ref{8.1}) and can be used to
simplify it. The corresponding transformation for matrix
(\ref{8.5}) can be represented as \be \label{8.6} g \to g'=U g
U^{-1} \ee where $U$ is a $3 \times 3$ matrix of the following
special form \be \label{8.7} U=\left( \ba{ccc}
1   &  0       &  0\\
b^1 &  K^{1}  &  K^{2}\\
b^2 &  -K^{2}  &  K^{1}  \ea \right). \ee

Up to equivalence transformations (\ref{8.6}), (\ref{8.7}) there
exist three matrices $g$, namely \be\label{8.28}
g_1=\lo\ba{ccc}0&0&0\\0&1&0\\0&0&1\ea\ro,\
g_2=\lo\ba{ccc}0&0&0\\1&0&0\\0&0&0\ea\ro,\
g_3=\lo\ba{ccc}0&0&0\\0&\al&-1\\0&1&\al\ea\ro.\ee

In accordance with (\ref{8.1}), (\ref{8.4}) the related symmetry
operator can be represented in one of the following forms \be
\label{8.11}\ba{l} X_1=\mu D-2(g_a)_{bc}\tilde u_c \p_{
u_b},\ X_2=e^{\lbd t}(g_a)_{bc}\tilde u_c \p_{ u_b} \ea
\ee or \be \label{8.12} X_3=e^{\lbd t+\om \cdot x} \p_{
u_2}. \ee Here $(g_a)_{bc}$ are elements of a chosen matrix
(\ref{8.28}), $b,c$ =0, 1, 2, $\tilde u=$ column $(1, u, v)$.

Formulae (\ref{8.11}) and (\ref{8.12}) give the principal
description of one-dimension algebras $A$ for equation (\ref{LG}).

To describe two-dimension algebras $\cal A$ we classify matrices
$g$ (\ref{8.5}) forming two-dimension Lie algebras. Choosing one
of the basis elements in the forms given in (\ref{8.28}) and the
other element in the general form (\ref{8.5}) we find that up to
equivalence transformations (\ref{8.6}) there exist three
two-dimension algebras of matrices $g$ \be\label{8.29}
A_{2,1}=\{g_1,g_3\}, \ A_{2,2}=\{g_2, g_4\},\ A_{2,3}=\{g_1
,g_2\}\ee  where $g_1, g_2, g_3$ are matrices (\ref{8.28}), and
$$g_4=\left( \ba{ccc}
0   &  0       &  0\\
0 &  0  &   0 \\
1 &  0  &  0  \ea \right). $$

Algebras $A_{2,1}$ and $A_{2,2}$ are Abelian while the basis
elements of $A_{2,3}$ satisfy $[g_1, g_2]= g_2$.

Using (\ref{8.29}) we easily find pairs of operators (\ref{4.2})
forming two-dimension Lie algebras. Denoting
\[
\hat e_\al=(e_\al)_{ab} {\tilde u}_b \p_{ u_a}, \quad
\al=1,2
\]
we represent them as follows: \be \label{8.17}\ba{l} < \mu D+ \hat
e_1+\nu  t \hat e_2, \hat e_2>,\ \ \ < \mu D+ \hat e_2+\nu  t \hat
e_1, \hat e_1>,\\<\mu D-\hat e_1,
         \nu D-\hat e_2>,\ \ \   <F_1 \hat e_1+ G_1 \hat e_2,\ F_2\hat e_1+G_2
 \hat e_2>\ea \ee for $e_1, e_2$ belonging to algebras $A_{2,1}$,
$A_{2,2}$, and \be \label{8.18} <\mu D- \hat e_1, \hat e_2>, \ \ \
 < \mu D+ \hat e_1+\nu  t \hat e_2, \hat e_2>\ee for
$e_1$ and $e_2$ belonging to $A_{2,3}$.

Here $\{F_1, G_1\}$ and $\{F_2,G_2\}$ are fundamental solutions of
the following system \be \label{8.20}  F_t=\lbd F+\nu G, \quad G_t
= \sigma F+\gm G \ee with arbitrary parameters $\lbd, \nu, \s,
\gamma$.

The list (\ref{8.17})-(\ref{8.18}) does not includes algebras of
the types \be\label{ag}< F\hat e,\ G\hat e>\ \ \text{and}\ \
<\mu D+e^{\nu t +\om \cdot x}\hat e,\ \
e^{\nu t +\om \cdot x}\hat e>\ee
(with $F,\ G$ satisfying (\ref{8.20}))
which are either incompatible with classifying equations (\ref{2.7}) or reduce to
one-dimension algebras. All the other two-dimension algebras $\cal A$
can be reduced to one the form given in (\ref{8.17}), (\ref{8.18})
using equivalence transformations (\ref{x.2}).

In analogous way we can find three- and four-dimensional algebras
of operators (\ref{4.2}).
Thus  we have two three-dimension algebras of matrices (\ref{8.5})
\be\ba{c}
A_{3,1}:\ \ e_1=g_1,\ e_2=g_2,\ e_3= g_4;\\
A_{3,2}:\ \ e_1=g_2,\ e_2=g_3,\ e_3=g_4\ea\label{n1}\ee
and the only four-dimension algebra:
\be A_{4,1}:\ \ e_1=g_1,\ e_2=g_3,\ e_3=\tilde g_4,\
e_4=g_2.\label{n2}\ee

Algebras (\ref{n1}) can be generalized to the following algebras $\cal A$
\be\label{8.24}\ba{l}<\mu D-2\hat e_1,\ \hat e_2,\  \hat e_3>,
\
<D+2\hat e_1+2\nu t\hat e_2,\  \hat e_2,\ \hat e_3>,\\
<D+2\hat e_1+2\nu t\hat e_3,\  \hat e_3,\ \hat e_2>,\ <\hat e_1,\
\ F_1 \hat e_2+G_1 \hat e_3,\ F_2\hat e_2+G_2 \hat e_3>\ea\ee
while (\ref{n1}) generate the following algebras:
\be\label{8.23}\ba{l}<\mu D-2\hat e_1,\  \hat e_2, \hat
e_3>, \
  <\hat e_1,  \ D+2\hat e_2+2\mu t \hat e_3,\ \hat e_3>.\ea\ee
  Finally, $A_{4,2}$
generates the following algebras $\cal A$
\be\ba{l}
<\mu D-2\hat e_1, \ \nu D-2\hat e_2, \ \hat e_3, \ \hat e_4>,\ \
<\hat e_1, \ \hat e_2, \ \hat e^{\mu t+\nu \cdot x}e_3,\ \hat
e^{\mu t+\nu \cdot x}e_4>.\ea\label{n3}\ee

The list (\ref{8.24})-(\ref{n3}) does not include algebras which
have subalgebras (\ref{ag}) (which are incompatible with classifying equations (\ref{4.3})).

\section{Classification results}

Using results presented in previous section we easily perform
group classification of equations (\ref{LG}). To do it we solve
the classifying equations (\ref{4.3}) with their known
coefficients $C^{ab}$ and $B^a$ which are defined comparing
(\ref{4.2}) with the found realizations of algebras $\cal A$.

If equation (\ref{LG}) admits one-dimensional algebra $\cal A$, i.e.,
one of algebras (\ref{8.11}), (\ref{8.12}) the related functions $f^1$ and $f^2$
have to satisfy  the corresponding determining equation (\ref{4.3} ) which
define $f^1$ and $f^2$  up to arbitrary
functions. In the case of two dimension algebras whose
realizations are given by relations (\ref{8.17}), (\ref{8.18}) we
 have a system of two determining equations corresponding to
two basis elements which usually define $f^1$ and $f^2$ up to
arbitrary parameters. In addition, we control the cases when
equation (\ref{LG}) admit extending symmetries $G_\mu$, $\widehat
G_\mu$ and $K$ (\ref{2.6}), i.e., when the found functions $f^1$
and $f^2$ satisfy conditions (\ref{4.8}), (\ref{4.10}) and
(\ref{!!}) respectively.

The next important step is to find additional equivalence transformations admitted by
equations (\ref{1.2}) with specified non-linearities $f^1$ and $f^2$. These transformations
are relatively easy calculated using the standard Lie algorithm and treating $f^1$ and $f^2$ as additional
variables.

We will not reproduce here the related routine calculations but
present their results in Tables 1-3, where non-linearities $f^1,
f^2$ and the related symmetries are specified. Additional
equivalence transformations are given here also.

The list of possible
AETs is present in the following formulae
\be\label{eqv}\ba{l}1.\ u\to\exp(\omega t)u, \ \ v\to \exp(\omega t)v,\\
2.\ u\to u\cos\omega t-v\sin\omega t,\ v\to v\cos\omega
t+u\sin\omega t,\\
3.\ u\to \exp(\omega t)u,\ v\to v+\omega \frac{t^2}{2},\\
4.\ u\to u+\omega t,\ v\to v,\\5.\ u\to u, \ v\to v+\omega t,\\
6.\ u\to \exp({\nu\omega t})\lo u\cos(\s\omega t)+v\sin(\s\omega t)\ro,\\
\ \ \ v\to \exp({\nu\omega t})\lo v\cos(\s\omega t)-u\sin(\s\omega t)\ro,
\\7.\ u\to \exp({2\omega t})\lo u\cos(\s\omega t^2)-v\sin(\s\omega t^2)\ro,\\
\ \ \ v\to \exp({2\omega t})\lo v\cos(\s\omega t^2)+u\sin(\s\omega t^2)\ro,\\
 8.\ u\to \exp({\lambda\omega t^2})\lo u\cos(2\omega t)+v\sin(2\omega t)\ro,\\
 \ \ \ v\to \exp({\lambda\omega t^2})\lo v\cos(2\omega t)-u\sin(2\omega t)\ro
 \ea\ee
where the Greek letters denote parameters whose values will be
specified in the tables. In contrast with (\ref{2.2}), (\ref{x.2}) these transformations
are valid only for equations with some special non-linearities $f^1, f^2$
specified in the Tables.

Table 1 presents non-equivalent equations which admit the main or the main
and extended symmetries.
The conditions for non-linearities which extend the symmetries are
specified in the fourth column. The additional equivalence transformations
are presented in the last (fifth) column.

In Table 2 the non-linearities are collected which correspond to the main
symmetries only. The additional equivalence transformations are specified
in the fifth column.

In Table 3 symmetries of a subclass of equations
(\ref{LG}) are specified. The additional equivalence transformations
are given in square brackets and placed in the fourth column.

In Tables 1-3 $G_\mu$, $\widehat G_\mu$ and $K$ are operators (\ref{2.6})
were $A$ is matrix of type $I$ (\ref{2.1}), $ \Psi_\mu(x)$ is an arbitrary
solutions of the Laplace
 equation:
 $$\Delta \Psi_\mu=
\mu\Psi_\mu.$$ The Greek letters denote arbitrary parameters,
moreover, up to equivalence transformations we restrict ourselves to
$\va=\pm 1$ and $\kappa=0, \pm 1$.

In Table 1 $\ba{l}R=\lo
u^2+v^2\ro^\frac12,\ z=\tan^{-1}\lo\frac{v}{u}\ro
\end{array}$, $F_1$ and $F_2$ are arbitrary functions of $R \exp(\mu z)$.

 \newpage

 \begin{center}
{\bf Table 1. Non-linearities and extendible symmetries for equations
(\ref{LG})}

\end{center}
\begin{tabular}{|l|l|l|l|l|}
\hline \text{No} & \text{Nonlinear terms}
&$
\begin{array}{l}
\text
{Main}\\\text{symmetries} \\
\end{array}$
&$
\begin{array}{l}
\text{Additional} \\
\text{symmetries}
\end{array}$&$
\ba{l}\text {AET}\\(\ref{eqv}) \ea$
\\
\hline 1. &$
\begin{array}{l}
f^1=uF_1+vF_2 \\
-\kappa z\left( \mu u+v\right),\ea$ &$
\begin{array}{l}
e^{\kappa t}\left( \mu R\partial_R-\p_z\ro \ea$&$
\begin{array}[t]{l}
\widehat{G}_\alpha ,\ \text{if }\\ \mu =a,
\kappa \neq 0; \ea$&$\ba{l}\\   2, \text{if}\ \mu\ea$\\
\cline{4-4} &$\ba{c}f^2=vF_1-uF_2\\+\kappa z\left(u- \mu v
\right)\ea$ & &$\ba{l} G_\alpha ,\ \text{if }\\ \mu =a,\kappa =0
\end{array}$&$\ba{l}=\kappa=0 \\ \ea$\\
\hline
 2. &$
\begin{array}{l}
f^1=e^{\nu z}R^\s (\lambda u-\mu v),\\f^2=e^{\nu z}R^\s (\lambda
v+\mu u)\ea$&$
\begin{array}{l}
\s  D-u\partial_u-v \partial_v,\\\nu D-u
\partial_v+v\partial_u \ea$&$\ba{l}
 G_\alpha  \text{ if }\nu =a\s\\ \text{\&}\ K \
\text{if}\ \s =\frac 4m
\end{array}
$&$\ba{c}\ \ \ 6\ea$\\
\hline
\end{tabular}

\vspace{2mm}

In Table 2 $F_1$ and $F_2$ are arbitrary functions whose arguments
are specified in Column 3, $R$ and $z$ have the same meaning as in Table 1.

 \begin{center}
{\bf Table 2. Non-linearities and non-extendible symmetries for
equations (\ref{LG})}

\end{center}
\begin{tabular}{|l|l|c|l|l|}
\hline \text{No} & \text{Nonlinear terms} &
\begin{tabular}{l}
Argu- \\
ments\\ of \
$F_\al$
\end{tabular}
&$
\begin{array}{l}
\text
{Symmetries}\\
\end{array}$
&$ \ba{l}\text {AET}\\(\ref{eqv}) \ea$
\\
\hline
 1. & $\begin{array}{l}
f^1=u^{\nu +1}F_{1,} \\
f^2=u^{\nu +1 }F_2
\end{array}$
& $\frac{u}{v}$ &$\begin{array}{l} \nu D-u{\partial_ u}-  v{
{\partial_ v}}
\end{array}$&$\ba{l}1\ \text{if}\\
 \nu=0\ea$
\\
\hline
 2. &$\ba{l} f^1=\beta u+F_1,\\ f^2=-\kappa u+F_2 \ea$&$ v
$&$ e^{(\beta +a\kappa
)t}\Psi _\kappa(x) \partial_u $&$\ba{c}4 \ \texttt{if}\\\kappa=\beta=0\ea$\\
\hline 3. &$
\begin{array}{l}
f^1=e^{\kappa v}F_1, \\
f^2=e^{\kappa v}F_2
\end{array}$
& $\ba{l}u\ea$  & $\kappa D-\partial_v$&  \\
\hline 4.&$\ba{l}f^1=u(F_1+\va\ln u),\\ f^2=v(F_2+\va\ln
u)\ea$&$\frac vu$&$e^{\va t}(u\p_u+v\p_v)$& $\ba{c}\ea$\\\hline
5.&$\ba{l}f^1=e^{\lambda z}(uF_1+vF_2),\\f^2=e^{\lambda
z}(vF_1-uF_2)\ea$& $Re^{\nu z}$&$\lambda D+\nu R\p_R-\p_z$& $\ba{c}
6, \ \s=1 \\ \texttt{if}\ \lambda=0\ea$\\\hline 6.
&$\ba{l}f^1=\kappa v^{\nu +1},\\ f^2=\beta v^{\nu +1}\ea $&&$\ba{l}
\nu D-u
\partial_u-v\partial_v,\\ \Psi_0
(x)\partial_u \ea$&$\ \ \ \ 4$\\
\hline 7. &$\ba{l} f^1=\kappa e^{v},\\  f^2=\beta e^{v} \ea$&&$
\ba{l}D-\partial_v,\  \Psi_0 (x)\partial_u\ea$&$\ \ \ \ 4$\\
\hline 8.&$ \ba{l} f^1=\mu \ln v,\\ f^2=\va \ln v \ea$&&$
\begin{array}{l}
\Psi_0 (x)\partial_u ,\\
D+u\partial_u+v\partial_v\\
{+}\left( (\mu-\va a )t\right.\\\left.-\frac
 {\va}{2 m}x^2\right)\partial_u
\end{array}
$&$\ \ \ \ 4$\\
\hline
9.&$\ba{l} f^1=\lambda,\\  f^2=\ln u \ea$&&$
\begin{array}{l}
 D+u\partial_u+v\partial_v+ t\partial_v,\\
\Psi _0(x) \partial_v
\end{array}
$&$\ \ \ 3, 5$\\
\hline
\end{tabular}

\vspace{3mm}

\newpage
\begin{center}
{\bf Table 3.  Symmetries of equations (\ref{LG}) with non-linearities
$f^1=(\mu u-\sigma v)\ln R+z(\lambda u- \nu v)$, $f^2=(\mu
v+\sigma u)\ln R+z(\lambda v+\nu u)$}
\end{center}
\begin{tabular}{|l|l|l|l|}
\hline
No & Conditions & Main symmetries & Additional   \\
& for coefficients &and AET (\ref{eqv})  & symmetries   \\ \hline $1$ &
$\ba{l}\lambda =0,\\\mu =\nu\ea$ & $e^{\mu t}\partial_ z,\
 e^{\mu t}\lo R\partial_ R+\sigma
t\partial_z\ro$ &
$\ba{l}
\hat{G}_\al \ \text {if}\  a\s=0,\\\mu \neq 0\ea$   \\
\cline{4-4} &   & [AET 7 if $\mu=0$] &$\ba{l}
  G_\al\
\text{if} \\ a=\nu =0,\s\neq 0\ea$   \\
\hline
$2$ & $\ba{l}\lambda =0,\\\mu \neq \nu,\ea$ & $\ba{l}e^{\nu
t}\partial_ z
,\\
e^{\mu t}\left( \sigma \partial_z+\left( \mu -\nu \right)
R\partial_ R\right) \ea$ &
$\ba{l}\hat{G}_\al\ \text{ if }\mu \neq 0\\  a\sigma =\nu -\mu ,\\
\text{or}\
a=0,\ \mu \neq 0\ea$\\
\cline{4-4} &  &[AET 6 if $\mu\nu=0$]  &$\ba{l} G_\al\ \text{if}\\
a\sigma =\nu
,\  \mu =0\ \ea$   \\
 \hline
 $3$ & $\ba{l}\delta=\frac14(\mu-\nu)^2\\+\lambda\s=0,\ea$ &
$\ba{l}X_3=e^{\omega _0t}\left( 2\lambda R\partial_
R+\left( \nu -\mu \right) \partial_ z \right),
\\2e^{\omega _0t}\partial_ z +tX_3\
\ea$ &$\ba{l} {\hat G}_\al\ \text{if} \ \om_0 \neq 0
\\ a(\mu -\nu
)=2\lambda \ea$   \\
\cline{4-4} &$\ba{l}\mu+\nu=2\omega_0\\
\lambda\neq 0  \ea$&$\ba{l} \texttt{[AET 6 if} \mu+\nu=0{]},\\
\texttt{{[}AET 8}\  \&\ 1 \texttt{ if }\ \mu=\nu=0{]}\ea$ &$\ba{l}
G_\al\ \text{if}\\ a\nu =-\lambda
,\ \omega_0=0 \ea$   \\
    \hline
    $4$ & $\ba{l}\lambda\neq 0,\ \delta=1,\\\om_\pm=\om_0\pm1\ea $ &
$\ba{l}e^{\omega _{+}t}\left( \lambda R\partial_ R+\left( \omega
_{+}-\mu \right) \partial_ z \right),\\
e^{\omega _{-}t}\left( \lambda R\partial_R+\left( \omega
_{-}-\mu \right)
\partial_ z \right) \ea$ &$\ba[t]{l}\widehat G_\al\ \text{if}\ \mu\nu\neq\lbd\s,\\
\lambda=a(\mu-\nu+a\s)\ea$   \\
\cline{4-4} & & $\ba{l}{[}\texttt{AET 6 } \texttt{if
}\mu\nu=\lambda\s, \\ \&\ 1 \ \texttt{if }\mu=\s=0 {]}\ea$
&$\ba{l}G_\al\
\text{if} \ \nu\mu=\lambda\s,\\
\lbd=a\mu\ea$
 \\
\hline $5$ & $\delta=-1$ & $\ba{l} \exp (\omega_0t)\left[2
\lambda\cos  tR\p_ R\right.\\ \left.+\left( \left( \nu-\mu\right)
\cos  t- 2\sin t\right)\p_ z \right],\\\exp
(\omega_0t)\left[2 \lambda\sin tR{\p_ R}\right.\\
\left.+\left( \left( \nu-\mu\right) \sin t+ 2\cos t\right){\p_ z}
\right] \ea $ & none   \\ \hline
\end{tabular}

\vspace{2mm}

\section{Discussion}

We perform group classification of generalized CLG equations
(\ref{LG}), i.e., find all non-equivalent equations of the
considered type and describe their symmetries. The obtained results
can be used  to construct exact solutions for equations which admit
sufficiently extended symmetries, using the standard Lie algorithm
\cite{olver}. The other application is to search for models with
{\it a priory} required symmetry, e.g., Galilei-invariance.

In Tables 1-3 all non-equivalent generalized CLG equations are
listed together with their symmetries and additional equivalence
transformations. More exactly, we specify here only extensions of
the basic symmetries (\ref{4.1}) and did not consider linear equations.

First we notice that the usual CGL equation appears as a particular case of
the classification procedure. The related non-linearities and symmetries are present in
Table 1, Item 2 when $\mu=0$. In addition to the basic symmetries (\ref{4.1}) this equation admits
the dilatation symmetry and symmetry $u\p_v-v\p_u$, which correspond to scaling of
dependent and independent variables and multiplying solutions of the CLG equation by a phase factor.

In accordance with our classification there exist eight types of generalized CLG
equations defined up to arbitrary functions $F_1$ and $F_2$
depending on variables indicated in the third column of Table 1,
Item 1 and Table 2, Items 1-5. Nonlinearities corresponding to the
most extended symmetries are collected in Table 1. In particular
there are nonlinearities corresponding to Galilei-invariant
equations (\ref{LG}) (refer to Table 1, Item 1 for $a=\mu,
\nu=0$):
$$
\begin{array}{l}
u_t-\Delta(au-v)=uF_1+vF_2, \\
v_t-\Delta(av+u)=vF_1-uF_2 \ea$$ where $F_1$ and $F_2$ are
arbitrary functions of $R{}e^{a z}$. This system can be rewritten
as a single equation for a complex function $W=u+iv$:
\be\label{exz1}W_t-(a+i)\Delta W={\cal F}W\ee where ${\cal
F}=F_1-iF_2$ is a complex function of real variable
$\xi=\ln|W|+az$ with $z$ being a phase of $W$.

The standard CLG equation does not belong to the class
(\ref{exz1}) and so is not Galilei invariant. On the other hand,
setting in (\ref{exz1}) $a=0$ we come to the Galilei-invariant
subclass of the NS equations (\ref{d6}).

The non-linearities enumerated in Table 3, Item 2 of Table 1 and
Items 6-9 of Table 2 are defined up to arbitrary parameters. The
most extended symmetry is indicated in Item 2 of Table 1 and
corresponds to the following equation for complex function
$W=u+iv$:
     \be\label{exz2}W_t-(a+i)\Delta W=\al \lo
e^{az}|W|\ro^{\rho}W\ee where $\al= \lambda+i\s$ is a complex
parameter and $\rho=\frac{4}{m}$.

In accordance with the above equation (\ref{exz2}) admits Lie
algebra of the Shr\"odinger group including operators $P_\mu,
J_{\mu\nu}$ (\ref{4.1}) and also generators of dilatation $D$,
Galilean transformations $G_\mu$ and conformal transformations $K$
(\ref{2.6}). Setting in (\ref{exz2}) $a=0$ we reduce (\ref{exz2})
to the very popular NS equation with  critical power $4/m$
non-linearity.

If $\rho\neq \frac{4}{m}$ then equation (\ref{exz2}) admit all
the above mentioned symmetries except the generator $K$ of conformal
transformations.

In general we indicate six classes of equations (\ref{LG}) which
admit symmetries $G_\mu$ and so are invariant with respect to
Galilei group. Namely, in addition to (\ref{exz1}), (\ref{exz2})
there are the following Galilei-invariant equations:
    \be\label{exz3}  iW_t+\Delta W=-\s W\ln|W|\ee
   which corresponds to Item 1, $\nu=0$ of Table 3, and
    \be\label{exz4}W_t-(a+i)\Delta W=cW(\ln |W|+az)\ee
where $c$ is a complex number equal to $i\s,\ \s(i-a)$ or
$\mu+i\s$ for versions indicated in Items 2, 3 or 4 of Table 3 correspondingly.

  Non-linearities collected in Table 2 correspond to equations
  (\ref{LG}) which admit the main and basic symmetries only.

 Thus we
present completed group classification of systems of
reaction-diffusion equations (\ref{1.2}) with square diffusion
matrix of type $I$ (\ref{2.1}).

The additional aim of this paper is to present an effective
approach for solving classifying equations (\ref{2.7}). It was
demonstrated in Sections 3 and 4 that the problem of group
classification of equations (\ref{1.2}) can be effectively reduced
to searching for the main symmetries (\ref{1.2}). We also make a
priori specification of these symmetries using the fact that they
should form a basis of a Lie algebra. The idea of such a
specification was proposed  in papers \cite{wint} and
\cite{zhdan1}.

In our following publications we use the approach described here
to classify reaction-diffusion equations with diagonal and
triangular diffusion matrix. In other words we plane to complete
classification of  equations (\ref{1.2}) with general diffusion
matrix whose non-equivalent versions are given by formulae
(\ref{2.1}).

\end{document}